\documentstyle{article}

\input epsf

\begin{document}

\baselineskip = 18pt

\thispagestyle{empty}
\rightline{LNF-96/058 (P)}
\rightline{hep-th/9611018 \vspace{1cm}}
\begin{center}
{\bf Gravitational interaction to one loop
in effective quantum gravity} \vspace{1.5cm} \\
A. Akhundov${}^{a}$
S. Bellucci${}^b$
A. Shiekh${}^{c}$\footnote{E-mail addresses: akhundov@azearn.ab.az,
bellucci@lnf.infn.it, shiekh@ictp.trieste.it} 
                    \vspace{1cm} \\
${}^a$Universit\"at-Gesamthochschule Siegen, D-57076 Siegen, Germany, and
Institute of Physics,\\
Azerbaijan Academy of Sciences,
pr. Azizbekova 33, AZ-370143 Baku, Azerbaijan\\
${}^b$INFN-Laboratori Nazionali di Frascati, P.O. Box 13, 00044 Frascati,
            Italy \\
${}^c$International Centre for Theoretical Physics, Strada Costiera 11,
P.O. Box 586, 34014 Trieste, Italy
\vspace{2.5cm} \\
{\bf Abstract} 
\end{center}
We carry out the first step of a program conceived, in order to build
a realistic model, having the particle
spectrum of the standard model and renormalized masses, interaction terms
and couplings, etc.
which include the class of quantum gravity corrections, obtained by
handling gravity as an effective theory. This provides an adequate picture
at low energies, i.e. much less than the scale of strong gravity (the Planck
mass). Hence our results are valid, irrespectively of any proposal for the
full quantum gravity as a fundamental theory. We consider only non-analytic
contributions to the one-loop scattering matrix
elements, which provide the dominant quantum effect at long distance.
These contributions are finite and independent from the
finite value of the renormalization counter terms of the effective lagrangian.
We calculate the interaction of two heavy scalar particles,
i.e. close to rest, due to the effective quantum gravity
to the one loop order and compare with similar results in the literature.
%. 
%This yields one loop quantum corrections to Newton's
%potential, which disagree with recent results obtained by Donoghue,
%although we do not dispute the validity of his approach, but just the details
%of his calculation.
\vfill
\begin{center}
November 1996
\end{center}
\setcounter{page}0
\renewcommand{\thefootnote}{\arabic{footnote}}
\setcounter{footnote}0
\newpage
%\rightline{\Huge Draft: 17th September 96}

\section{Introduction}

A longstanding puzzle in quantum physics is how to marry the description
of gravity with the field theory of elementary particles. Introducing
an effective low-energy theory for processes with a typical energy
less the Planck mass, i.e. with $|q^2|\ll M_{Planck}^2\simeq 10^{38}$ GeV$^2$,
was suggested by Weinberg \cite{wein1}. This effective theory could be
modeled after the example of Chiral Perturbation Theory \cite{gl}
for QCD. The important difference to keep in mind, when considering the
two cases, is that the analogous of QCD, i.e. a fundamental theory of gravity,
is still lacking.

At very low energy, the theory is provided by general relativity. The natural
question to address is then the following: what about quantum corrections?
There is a problem that arises immediately when trying to answer this
question, namely that the $O(p^4)$ constants needed for renormalization
are unknown. Indeed, one does not have available experimentally
measured quantities, such as the graviton-graviton scattering lengths,
in order to determine the finite value of the renormalized counter terms
of the effective lagrangian.
Donoghue showed first in \cite{d1} how to go about this problem and we will
handle the matter in a similar way, which explains why we will never treat
divergences in the calculation of quantum effects, but only terms with a
non-analytic behavior, such as $\ln(-q^2)$ and 1/$\sqrt{-q^2}$.
We essentially restrict ourselves to
calculating only these
non-analytic contributions to the scattering matrix elements.

The program consists then in building a realistic model with the particle
spectrum of the standard model and renormalized masses, interaction couplings,
etc. which include the class of quantum gravity corrections, obtained by
handling gravity as an effective theory. This provides an adequate picture
at low energies, i.e. much less than the scale of strong gravity (the Planck
mass). In this Letter, we start carrying out this program. We begin
with scalar fields coupled to quantum gravity and deal with gravity as an
effective theory. We calculate the interaction of two heavy scalar particles,
i.e. with $m_{1,2}\gg\sqrt{-q^2}$, due to the effective quantum gravity
to the one loop order. This yields one loop quantum corrections to Newton's
potential, which disagree with Donoghue's results of \cite{d1,d2}.
Notice, however, that we do not dispute the validity of the approach, but
just the details of the calculation and the results
in \cite{d1,d2}. Also, we wish to stress
that our results hold their validity at low energy, not needing any definite
proposal for the complete quantum gravity theory.\footnote{
A tentative approach is made in \cite{andy} to quantizing gravity
with a modified renormalization scheme, and is
illustrated for the example of a massive scalar field
with gravity.}

The outline of this Letter is as follows.

We begin in Section 2 with giving the most general form of the corrections
to the Newtonian potential, including the lowest-order correction terms. We also
present the Feynman rules in 4 dimensions, discussing the (lack) of
importance of the regularization used for the lagrangian. The calculation
of the vertex correction is the object of Section 3, where it is described
in some detail, in order to make the comparison with
Donoghue's calculation and point out the source of the disagreement with it.
We also discuss briefly the strategy of our calculation which involves
manipulations with the computer. The complete one loop correction to
Newton's potential, obtained by adding the contribution of the
graviton vacuum polarization, is presented in Section 4. We close the Letter
with some concluding remarks, a summary of the results obtained and
the outlook of possible developments of this investigation.

\section{The lowest-order corrections to the Newtonian potential}

Our notations and conventions on the metric tensor, the gauge-fixed
gravitational action, etc. are the same as in \cite{d1,d2}.
On the grounds of dimensional analysis alone
one can anticipate the form of the lowest-order corrections
to the Newtonian potential in configuration space \cite{d2}:

\begin{equation}
V(r) =
-{{Gm_1m_2} \over r}
\left(
{1 + \alpha{{G(m_1 + m_2)} \over {rc^2}}
+ \beta{{G \hbar} \over {r^2 c^3}} + \ldots}
\right) \quad ,
\end{equation}

\noindent which includes the lowest-order relativistic correction,
and the lowest-order quantum correction (also relativistic).
These power law corrections come from momentum space
(where we normally calculate), via the Fourier transforms
\cite{d1,d2}:

\begin{equation}
\int {{{d^3q} \over {(2\pi)^3}} e^{-i\vec q.\vec r}}
{1 \over {\sqrt{{\vec q \,}^2}}}
=
{1 \over {2\pi^2 r^2}}
\end{equation}
and

\begin{equation}
\int {{{d^3q} \over {(2\pi)^3}} e^{-i\vec q.\vec r}}
\ln \left( { {\vec q \,}^2 \over \mu^2} \right) 
=
-{1 \over {2\pi^2r^3}} \quad .
\end{equation}

Notice that the Fourier transform in (3) seems
badly defined, since it is affected by a $\delta^3(r)$ ambiguity.
Indeed the $\delta$-terms correspond to point-like interactions,
which we are not interested in. We are instead
interested in the leading quantum effects, i.e. the non-analytic
contributions, which dominate at long distance
(see the discussion after eq. (4) of \cite{d1}). Since these dominant
(at large $r$) effects are insensitive to the addition of point-like
interactions to the lagrangian, the $\delta^3(r)$ ambiguity in (3)
is completely irrelevant for our calculation.

The importance of these transforms, is that they are
from non-analytic terms in momentum space and so cannot
be renormalized into the original Lagrangian, and as such
one might anticipate that they are of finite magnitude.
Because of this,
the problem of renormalizing quantum gravity is put off.

Now there is a further magic in the second transform,
where the scale $\mu$ does not feed through to
configuration space. So this particular calculation
is free of the ambiguities normally intrinsic to
quantum field theory.

\subsection{The lowest-order Feynman rules}

In gravity there are an infinite number of Feynman rules,
although this does not cause a problem, since only a finite
number are used when working to any finite order.
The lowest-order ones we list as \cite{d2}:

$\bullet$ The graviton propagator:
\vglue .5cm
\hskip .8cm
\epsfbox{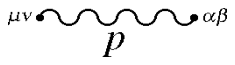}
\vglue -1.5cm

\begin{equation}
iD_{\mu \nu \alpha \beta}(p)
={i \over {p^2 + i\varepsilon}} P_{\mu \nu \alpha \beta}
\end{equation}

\rightline{ \small
where
$
P_{\mu \nu \alpha \beta}
\equiv {\textstyle{1 \over 2}}(\eta _{\mu \alpha}\eta _{\nu \beta}
+ \eta _{\mu \beta} \eta _{\nu \alpha}
-\eta _{\mu \nu}\eta _{\alpha \beta})
$
}

\vglue .5cm
$\bullet$ The scalar propagator:
\vglue .5cm
\hskip 1cm
\epsfbox{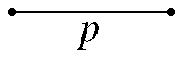}
\vglue -1.5cm

\begin{equation}
{i \over {p^2 - m^2 + i\varepsilon}}
\end{equation}

\vglue .5cm
$\bullet$ The scalar-graviton vertex:
\vglue 1.5cm
\hskip 1cm
\epsfbox{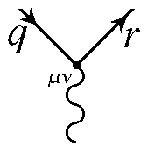}
\vglue -1.5cm

\begin{equation}
\tau_{\mu \nu}(q,r,m)
=
-{\textstyle{1 \over 2}}i\kappa 
\left(
{q_\mu r_\nu  + r_\mu q_\nu -\eta_{\mu \nu}(q.r - m^2)}
\right)
\end{equation}

\vglue .5cm
$\bullet$ The scalar-two graviton vertex:
\vglue 1.5cm
\hskip 1cm
\epsfbox{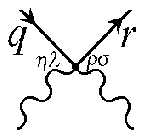}
\vglue -2.0cm

\begin{equation}
\tau_{\eta \lambda \rho \sigma}(q,r,m)
=\textstyle{1 \over 2} i\kappa^2
\left( 
\begin{array}{lll}
&I_{\eta \lambda \alpha \delta} {I^\delta}_{\beta \rho \sigma}
(q^\alpha r^\beta  + r^\alpha q^\beta ) \\[3mm]
-&{\textstyle{1 \over 2}}(\eta _{\eta \lambda} I_{\rho \sigma \alpha \beta}
+ \eta _{\rho \sigma} I_{\eta \lambda \alpha \beta})r^\alpha q^\beta \\[3mm]
-&{\textstyle{1 \over 2}}(I_{\eta \lambda \rho \sigma}
-{\textstyle{1 \over 2}}\eta _{\eta \lambda}\eta _{\rho \sigma})
(q.r-m^2)
\end{array}
\right)
\end{equation}

\rightline{ \small
where
$
I_{\mu \nu \alpha \beta}
\equiv {\textstyle{1 \over 2}}
(\eta_{\mu \alpha}\eta_{\nu \beta}
+ \eta_{\mu \beta}\eta_{\nu \alpha})
$
}

\vglue .5cm
$\bullet$ The three graviton vertex:
\vglue 1.5cm
\hskip 1cm
\epsfbox{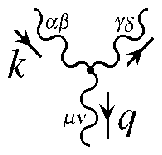}
\vglue -1.5cm

\begin{equation}
\begin{array}{lllllll}
\hskip 5cm {\tau^{\mu \nu}}_{\alpha \beta \gamma \delta}(k,q) = \\[7mm]
{\textstyle{1 \over 2}}i\kappa
\left( 
\begin{array}{lll}
&P_{\alpha \beta \gamma \delta}\left( {k^\mu k^\nu  + (k-q)^\mu (k-q)^\nu 
+ q^\mu q^\nu -{\textstyle{3 \over 2}}\eta ^{\mu \nu}q^2} \right) \\[3mm]
+ &2q_\lambda q_\sigma \left( { {I^{\lambda \sigma}}_{\alpha \beta} 
{I^{\mu \nu}}_{\gamma \delta}
+ {I^{\lambda \sigma}}_{\gamma \delta} {I^{\mu \nu}}_{\alpha \beta}
-{I^{\lambda \mu}}_{\alpha \beta} {I^{\sigma \nu}}_{\gamma \delta}
-{I^{\sigma \nu}}_{\alpha \beta} 
{I^{\lambda \mu}}_{\gamma \delta}} \right) \\[3mm] 
+ &q_\lambda q^\mu \left( {\eta _{\alpha \beta}
{I^{\lambda \nu}}_{\gamma \delta} + \eta _{\gamma \delta}
{I^{\lambda \nu}}_{\alpha \beta}}\right)
+ q_\lambda q^\nu \left({\eta _{\alpha \beta} 
{I^{\lambda \mu}}_{\gamma \delta} + \eta _{\gamma \delta} 
{I^{\lambda \mu}}_{\alpha \beta}} \right)\\[3mm] 
-&q^2\left( {\eta _{\alpha \beta} 
{I^{\mu \nu}}_{\gamma \delta} + \eta _{\gamma \delta}
{I^{\mu \nu}}_{\alpha \beta}}\right)-\eta ^{\mu \nu}
q^\lambda q^\sigma \left( {\eta _{\alpha \beta}
I_{\gamma \delta \lambda \sigma} + \eta _{\gamma\delta} 
I_{\alpha \beta \lambda \sigma}} \right) \\[3mm]
+ &2q^\lambda \left( { {I^{\sigma \nu}}_{\alpha \beta} 
I_{\gamma \delta \lambda \sigma}(k-q)^\mu 
+ {I^{\sigma \mu}}_{\alpha \beta}  I_{\gamma \delta \lambda \sigma}(k-q)^\nu
-{I^{\sigma \nu}}_{\gamma \delta} I_{\alpha \beta \lambda \sigma}k^\mu 
-{I^{\sigma \mu}}_{\gamma \delta} I_{\alpha \beta \lambda \sigma}k^\nu} \right) \\[3mm]
+ &q^2\left( { {I^{\sigma \mu}}_{\alpha \beta} {I_{\gamma \delta \sigma}}^\nu 
+ {I^{\sigma \mu}}_{\gamma \delta} 
{I_{\alpha \beta \sigma}}^\nu} \right) + \eta ^{\mu \nu}q^\lambda q_\sigma 
\left( { {I^{\rho \sigma}}_{\gamma \delta}I_{\alpha \beta \lambda \rho}
+ {I^{\rho \sigma}}_{\alpha \beta} I_{\gamma \delta \lambda \rho}} \right)
\\[3mm] 
+ &\left( {k^2 + (k-q)^2} \right)\left( { {I^{\sigma \mu}}_{\alpha \beta}
{I_{\gamma \delta \sigma}}^\nu + {I^{\sigma \nu}}_{\alpha \beta}
{I_{\gamma \delta \sigma}}^\mu -{\textstyle{1 \over 2}}\eta ^{\mu \nu}
P_{\alpha \beta \gamma \delta}} \right) \\[3mm]
-&\left( {k^2\eta _{\gamma \delta}
{I^{\mu \nu}}_{\alpha \beta} + (k-q)^2\eta _{\alpha \beta}
{I^{\mu \nu}}_{\gamma \delta}}
\right)
\end{array}
\right)
\end{array}
\end{equation}

%\marginpar{\em Small typo in the Feynman rule fixed here.}

These Feynman rules are presented in 4 dimensions, which
would be a problem if we anticipate using dimensional regularization
later. However, since
we are not dealing with infinite parts at all in this particular
calculation, it will not make a difference. Further, there
does exist, in the form of operator regularization \cite{reg}, a
method not departing from 4 dimensions, and the presented
rules are then complete if one were to proceed using this
technique. This technique is fundamentally an analytic
continuation, and a generalization of the one-loop zeta
function technique. One regulates the divergent piece using:

\begin{equation}
\Omega^{-m} = 
\lim \limits_{\varepsilon \to 0} {d^n \over d \varepsilon^n}
\left( (1 + \alpha_1 \varepsilon + \ldots  + \alpha_n \varepsilon^n)
{\varepsilon^n \over n!}
\Omega^{-\varepsilon - m}
\right)
\end{equation}
\rightline{\small \it (the alphas being ambiguous)}

We have taken the liberty to write down the most general
form of this equation, which is normally presented in a
more specialised form. We do this, as the difference is
critical  when it comes to quantum gravity.

\section{Vertex corrections}

Only two one-loop vertex contractions contribute
non-analytic terms, and they are:

\vglue .5cm
$\bullet$ The graviton loop vertex correction (1a):
\vglue 1.0cm
\hskip 1cm
\epsfbox{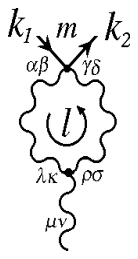}
\vglue -1.5cm

\begin{equation}
M_{1a}^{\mu \nu}\equiv iP^{\alpha \beta \lambda \kappa}
iP^{\gamma \delta \rho \sigma}\tau _{\alpha \beta \gamma \delta
}(k_1,k_2,m)\int {{{d^4l} \over {(2\pi )^4}}}
{{ {\tau ^{\mu \nu}}_{\lambda \kappa \rho \sigma}(l,q)} 
\over {l^2\left( {l-q} \right)^2}}
\end{equation}

\rightline{ \small
where $q \equiv k_1 - k_2$.
}

After performing the tensor algebra on a computer, using the
Ricci package \cite{lee} under the Mathematica \cite{m} program, one
arrives at:

%\marginpar{\em Not sure we could have done it
%without Ricci. Would like to acknowledge Lee properly.}

\begin{equation}
M_{1a}^{\mu \nu}
={{\kappa ^3} \over 4}
\int{{d^4l} \over {(2\pi)^4}}
{\left( 
\begin{array}{lll}
&\left( {2(k_1^\mu k_2^\nu + k_1^\nu k_2^\mu )
- g^{\mu \nu}k_1.k_2} \right)l^2 \\[3mm]
+ &\left( {5m^2-3k_1.k_2} \right)
(2l^\mu l^\nu -q^\mu l^\nu - q^\nu l^\mu ) \\[3mm]
+ &{\textstyle{g^{\mu \nu}} \over \textstyle 2}(3k_1.k_2-7m^2) q^2
+ 3(2m^2 - k_1.k_2)q^\mu q^\nu  \\[3mm]
\end{array}
\right)
\over {l^2 \left( {l-q} \right)^2}} \quad .
\end{equation}

Using the table of integrals \cite{d2} for the non-analytic
contribution, yields:

\begin{equation}
M_{1a}^{\mu \nu}
={{\kappa ^3} \over 4}{i \over {32\pi ^2}}
\left({-{L \over 3}} \right) 
\left( 
{26 m^2 - 12 k_1.k_2}
\right)
(q^\mu q^\nu - g^{\mu \nu}q^2)
\end{equation}

\rightline{ \small
where $L \equiv \ln ( -q^2/ \mu^2 )$.
}
This would agree with Donoghue if we were to drop the
$k_1.k_2$ term. Finally, going on shell using 
$k_1.k_2 = m^2 - {q^2 / 2}$ yields:

\begin{equation}
M_{1a}^{\mu \nu}
= -{{i\kappa ^3} \over {64\pi ^2}}{{7m^2+3q^2} \over 3}
(q^\mu q^\nu - g^{\mu \nu} q^2) L \quad .
\end{equation}
Hence onto the form factors:\footnote{Henceforth we neglect further
non-analytic
terms containing higher powers of $q^2$/$m^2$, since they do not contribute
to the leading physical effect we calculate, i.e. the leading quantum
corrections to the gravitational interaction of two {\it heavy} particles.}

\begin{equation}
V_{1a}^{\mu \nu}
= {{\kappa ^2} \over {32\pi^2}}{{7m^2} \over 3}
(q^\mu q^\nu - g^{\mu \nu} q^2) L \quad .
\end{equation}

\vglue .5cm
$\bullet$ The graviton-scalar loop vertex correction (1b):
\vglue 0.5cm
\hskip 1cm
\epsfbox{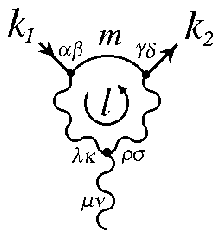}
\vglue -0.5cm

\begin{equation}
M_{1b}^{\mu \nu}=
iP^{\alpha \beta \lambda \kappa}iP^{\gamma \delta \rho \sigma}
i\int {{{d^4l} \over {(2\pi)^4}}}
{{\tau _{\alpha \beta}(k_1,k_1-l,m) \
\tau _{\gamma \delta}(k_1-l,k_2,m) \
{\tau^{\mu \nu}}_{\lambda \kappa \rho \sigma}(l,q)} \over 
{l^2\left( {l-q} \right)^2\left( {(l-k_1)^2-m^2} \right)}}\quad .
\end{equation}

Doing the algebra on the computer, using the integral tables
\cite{d2}, going on shell, and working to leading order in $q^2$/$m^2$,
yields:

\begin{equation}
M_{1b}^{\mu \nu}=
{{i\kappa^3} \over {64\pi^2}}\left( {{{q^2} \over 16}(20L + S)
\left( {k_1^\mu k_2^\nu + k_1^\nu k_2^\mu + q^2{{g^{\mu \nu}} \over 2}} 
\right) + {{7m^2} \over 24}(8L + 3S)(q^\mu q^\nu - g^{\mu \nu}q^2)}
\right)
\end{equation}

\rightline{ \small
where $S \equiv \pi^2 m / \sqrt{-q^2}$.
}
Hence onto the form factors

\begin{equation}
V_{1b}^{\mu \nu} =
-{{\kappa^2} \over {32\pi ^2}}
\left[
{{{q^2} \over 16}(20L + S)
\left(
{k_1^\mu k_2^\nu  + k_1^\nu k_2^\mu  + q^2{{g^{\mu \nu}} \over 2}} 
\right) 
+ {{7m^2} \over 24}(8L + 3S)(q^\mu q^\nu -g^{\mu \nu}q^2)}
\right]
\end{equation}

Adding (17) with (14) and including also the lowest-order (tree-level)
contribution, which can be obtained from (6)
\begin{equation}
V_0^{\mu\nu}=
{k_1^\mu k_2^\nu  + k_1^\nu k_2^\mu  + q^2{{g^{\mu \nu}} \over 2}} \quad ,
\end{equation}
yields
\begin{eqnarray}
V^{\mu\nu} &=& V_0^{\mu\nu}+V_{1a}^{\mu\nu}+V_{1b}^{\mu\nu}\nonumber \\
&=&
\left(1-{{\kappa^2} \over {32\pi ^2}}{{q^2} \over 16}(20L + S)\right)
\left(
{k_1^\mu k_2^\nu  + k_1^\nu k_2^\mu  + q^2{{g^{\mu \nu}} \over 2}} 
\right)\nonumber \\
&-& {{\kappa^2} \over {32\pi ^2}}
 {{7m^2} \over 8}S(q^\mu q^\nu -g^{\mu \nu}q^2)
\quad .
\end{eqnarray}
This yields the following contribution to the gravitational interaction
of two heavy particles (i.e. with $m_{1,2}^2\gg |q^2|$):
\begin{equation}
-{{\kappa ^2} \over 4}
{1 \over {2m_1}}V_{\mu \nu}^{(1)}
\left[ {iD^{\mu \nu \alpha \beta}} \right]
V_{\mu \nu}^{(2)}{1 \over {2m_2}}=-{{\kappa ^2} \over 4}\frac{m_1m_2}{2}i
\left[ {\frac{1}{q^2}-\frac{\kappa^2}{32\pi^2}
\left( \frac{5}{2}L+\frac{\pi^2 (m_1+m_2)}{2\sqrt{-q^2}}\right)} \right]
\quad .\label{vertex}
\end{equation}

In general, the Feynman rules are large and the tensor
algebra immense. Much of the calculational simplicity is
restored by submitting this part of the complexity to the
computer. However, the intermediate results can be so
extensive that even a super-computer can choke without help.
For example, imagine one had the contraction of three tensors:
$
\alpha^{\mu \nu} \beta^{\rho \sigma}
\gamma_{\mu \nu \rho \sigma}
$
each of which consists of many terms. Then the
computer, in trying to contract out the indices, tends to
expand out the entire expression, which can easily lead to
thousands of terms in the intermediate expression, and so
overpower the computers memory. The resolution lies in asking
the computer to initially expand out only $\alpha$ for example:
$
(\alpha_1^{\mu \nu} + \alpha_2^{\mu \nu} + \ldots)
\beta^{\rho \sigma} \gamma_{\mu \nu \rho \sigma}
$.
In this way the computer is presented with several terms that
can each be contracted separately. This seemingly innocuous
move can make all the difference between the machine being
able to perform the calculation or not. It is fine details
like this that in practice can occupy much of the investigators
time.

%\marginpar{\em avoided discussing Mathematica's {\tt Release}
%and \tt{Hold} commands.}

\section{Putting it all together}

{}From the counter Lagrangian (5.24) of 't Hooft and Veltman \cite{tv}, working
to second order in the quantum fluctuations $h$ of the gravitational
field and using eqs. (3.31), (3.33) of \cite{tv}, yields
the following expression for the contribution of the graviton plus ghost
vacuum-polarization Feynman-diagrams\footnote{As observed in \cite{d1},
the use of dimensionally
regularized matrix elements, together with the masslessness of the particles
running in the loop, allow
us to read off the $L$-term from the divergent
contribution due to the vacuum polarization.}:

\begin{equation}
\begin{array}{lll}
\Pi_{\alpha \beta \gamma \delta}= \\
-{\textstyle{\kappa ^2} \over \textstyle{16 \pi^2}} 
\left\{
\begin{array}{lll}
&{21 \over 120} q^4 I_{\alpha \beta \gamma \delta}
+ {23 \over 120} q^4 \eta _{\alpha \beta} \eta_{\gamma \delta}
-{23 \over 120} q^2
\left( 
{\eta_{\alpha \beta} q_\gamma q_\delta   + \eta_{\gamma \delta}
q_\alpha q_\beta}
\right) \\[3mm]
-&{21 \over 240} q^2 
\left( 
\eta_{\beta \gamma} q_\alpha q_\delta 
+ \eta_{\beta \delta} q_\alpha q_\gamma 
+ \eta_{\alpha \delta} q_\beta q_\gamma 
+ \eta_{\alpha \gamma} q_\beta q_\delta
\right)
+ {11 \over 30} q_\alpha q_\beta q_\gamma q_\delta
\end{array}
\right\}
L + (\textstyle{analytic~terms})
\end{array}
\quad ,
\end{equation}

\noindent where Donoghue's typographical errors have
been corrected. This contributes to the two-scalars
interaction due to gravitation, as follows:
\begin{eqnarray}
-{{\kappa ^2} \over 4}
{1 \over {2m_1}}V_{\mu \nu}^{(1)}
\left[ {iD^{\mu\nu\rho\sigma}i
\Pi_{\rho\sigma\eta\lambda}iD^{\eta\lambda\alpha\beta}} \right]
V_{\mu \nu}^{(2)}{1 \over {2m_2}}=-\frac{\kappa^2}{4}m_1m_2i
\left( \frac{\kappa^2}{32\pi^2}\right)
\left( \frac{43}{120}\right) L
\quad . \label{pol}
\end{eqnarray}
Here we neglected terms of order higher than one loop.
 
Collecting together the one-loop contributions due to the vertex
correction (\ref{vertex}) and the vacuum polarization (\ref{pol}),
we get, after taking the non-relativistic limit $p_{\mu}=(m,0,0,0)$,
$q=(0,{\vec q})$, for the interaction between two scalars

\begin{eqnarray}
-{{\kappa ^2} \over 4}
{1 \over {2m_1}}V_{\mu \nu}^{(1)}
\left[ iD^{\mu\nu\alpha\beta}+iD^{\mu\nu\rho\sigma}i\right.
\left. \Pi_{\rho\sigma\eta\lambda}iD^{\eta\lambda\alpha\beta} \right]
V_{\mu \nu}^{(2)}{1 \over {2m_2}} &=& 
4\pi Gm_1m_2i
\left[ \frac{1}{{\vec q \,}^2} \right. \nonumber \\
&+& \left. \frac{\kappa^2}{32\pi^2}
\left( \frac{107}{60}L+\frac{\pi^2(m_1+m_2)}{2\sqrt{{\vec q \,}^2}}\right)
\right]
\end{eqnarray}
with $32\pi G=\kappa^2$. This goes over to co-ordinate space as:

\begin{equation}
V(r) =
-{{G m_1 m_2} \over r}
\left[
{1 + {{G(m_1 + m_2)} \over {r c^2}} - 
\frac{107}{30\pi^2} {{G\hbar} \over {r^2 c^3}} + \ldots}
\right] \quad ,\label{four}
\end{equation}
which agrees with the form anticipated in (1), for the leading corrections
to the gravitational interaction of two heavy scalars. The quantum correction
agrees with \cite{d1,d2} only in size, but not in the details. Notice that
the sign of the second term in the r.h.s. of (\ref{four}) is opposite to
Donoghue's result.

As expected, the quantum corrections become large at
\begin{equation}
r_0\simeq\sqrt{\frac{G\hbar}{c^3}}=\frac{\hbar c}{M_{Pl}} \quad ,
\end{equation}
however in this regime the effective lagrangian approach breaks down.
Notice also that the counter terms (which are by definition short-range
effects) are expected to contribute only constant terms in (23) (to the
one-loop order). The latter reflect themselves in a $\delta^3 (r)$ contribution
to (24), with an unknown coefficient. It is only when the counter terms
are inserted as vertices in loop diagrams (i.e. at the two-loop level), that
they can contribute to a long-range effect, hence to (23).
Finally, at some scale $\mu_0$ of the order of (a
fraction of) the Planck mass, the constant contributions of the
counter terms in (23), which depend
on $\mu$ but not on $q$, will have the same size as the $L$-term.
Once again, there are, in this energy range, very large corrections to
the one-loop result, coming from higher-order terms in the perturbative
expansion, which then cannot be trusted any longer.

\section{Conclusion and outlook}

In this Letter we started an investigation aimed at obtaining
a class of the quantum corrections to the theory describing elementary
particles, e.g. the standard model, due to gravitational effects. Following
the pioneering work of Weinberg \cite{wein1}, and stimulated by more recent
seminal papers \cite{d1,d2}, we choose to treat quantum gravity within
the framework of effective field theories. This procedure yields results that
are valid at low energy, with respect to the Planck mass. We focus on
a theory of scalar fields coupled to the effective quantum gravity and
carry out the calculation of the one-loop corrections to the gravitational
interaction of two heavy scalars, i.e. close to rest. This requires
computing, to the one-loop order, the graviton plus ghost vacuum polarization
diagrams, as well as the vertex corrections.

The one-loop corrections are
divergent and require the introduction of appropriate renormalizing
counter terms in the effective lagrangian to $O(p^4)$.
Owing to obvious difficulties
in the determination of the finite value of the $O(p^4)$ constants,
e.g. relating them to quantities accessible to experiments,
the purely analytic terms are not included in the
calculation of the quantum corrections \cite{d2}.
Notice that this sector of quantum corrections
is ambiguous, since the purely analytic contributions involve the unknown
values of the $O(p^4)$ constants.
Hence we calculate that class of $O(p^4)$ corrections yielding effects
with a non-analytic behaviour, which are finite and do not depend
on the finite value of the renormalized counter terms in the effective
lagrangian. This class provides the leading quantum corrections, namely
the dominant ones for large $r$ \cite{d1}.

We have calculated
the non-analytic contribution of the vacuum-polarization diagrams to
the gravitational scalar-scalar interaction, as in \cite{d1,d2},
starting from the divergent counter terms of \cite{tv}. We performed the
calculation of the vertex corrections, as well. They have been calculated
before by Donoghue, but we do not agree with his results in \cite{d1,d2}.
We have checked our calculation, given their complexity and length, with
the use of computer programs
for algebraic manipulations, based on a package which proved very helpful
in dealing with
tensors in differential geometry
\cite{lee}, and believe our results are correct.
The explicit result for the class of quantum corrections to one loop cannot
be measured in experiments, because it is too small.\footnote{The experiments
searching for violations of the equivalence principle and deviations
from Newton's inverse square law constrain the mass of the Higgs-like
boson advocated in extended supergravity theories \cite{bf1}. These
high-precision experiments yield bounds for the antigravity fields appearing
in $N=2,8$ supergravity \cite{bf2}.}
Nonetheless,
it is not arbitrary, rather it follows from the low-energy theory.
Whatever the fundamental theory describing the quantum behaviour
of gravity, our result must hold its validity at low energies.

Finally, a calculation of the beta functions and anomalous dimensions
of a scalar field theory coupled to gravity on a flat background
appeared recently \cite{p}. The authors of \cite{p} treat Einstein's theory
as an effective field theory. It would be interesting to see if their
calculation is relevant, in connection with the effective lagrangian
approach to quantum gravity used to obtain our result.
\vspace{0.5cm}\\

\noindent{\large\bf Acknowledgments}

We would like to acknowledge
the scientific computer section of ICTP for making hours available
on their mainframes. We are grateful to J.F. Donoghue for discussions.
A.A. would like to thank the International Centre for
Theoretical Physics, INFN-Laboratori Nazionali di Frascati and 
Technische Hochschule Darmstadt for the hospitality 
extended to him during the course of this work and 
for the financial support and the INTAS for 
the financial support from the project INTAS-93-744. 
S.B. wishes to thank R. Percacci at SISSA for hospitality
when this work was undertaken.

\end{document}